# Investigation of Self-Collimation Properties in Non-Linear Low Symmetry Photonic Crystal Structures

Ozgur Onder Karakilinc
*Pamukkale University*
*Faculty of Engineering*
Electrical and Electronics Engineering
Denizli, Turkiye
okarakilinc@pau.edu.tr

*Abstract*— **In this study, the self-collimation properties of the low-symmetry hexagonal photonic crystal structure consisting of Kerr-type nonlinear LiNbO3 materials were extracted, and its transmission properties were examined according to the nonlinear index change.**

## I. Introduction

Photonic crystals (PCs) are periodic optical structures that create photonic bandgaps, prohibiting the propagation of certain wavelengths of light. Mathematical calculations in photonic crystals reveal complex interactions between light and periodic structures, leading to phenomena such as bandgaps, negative refraction, nonlinearity, and bistability [1]. Understanding these relationships through rigorous calculations (in both frequency and time domains) is essential for designing advanced photonic devices and applications in areas like telecommunications and optical sensing [1-5].

Nonlinear (NL) PCs, which exhibit a refractive index dependent on the intensity of the incident light, create conditions for enhanced light-matter interactions, enabling the control and manipulation of light propagation within the crystal. When Kerr nonlinearity is introduced, the refractive index changes proportionally to the light's electric field intensity, leading to phenomena such as soliton formation and modulational instability, which are critical for applications in all-optical switching and signal processing [6].

Self-collimation (SC), a key effect in photonic crystals, refers to the ability of certain PCs to direct light beams through the crystal without divergence. This occurs when the equi-frequency contours (EFCs) in the photonic band structure are nearly flat. A flat EFC implies that the group velocity of light is consistent across different wavevectors, resulting in the self-collimation of light. In practical designs, the self-collimation effect is often achieved by tailoring the geometry and refractive index contrast of the photonic crystal to flatten the EFCs in the desired direction [2,7,8]. This effect is particularly useful in optical circuits where precise control over light propagation is essential. Self-collimation is utilized in creating compact photonic devices, such as waveguides, splitters, and interferometers, where minimizing beam divergence is crucial for maintaining signal integrity [9-13].

When combined with Kerr-type nonlinearity, the self-collimation effect can be dynamically tuned, allowing for precise control over the beam's propagation path by adjusting the light intensity. This tunability is particularly beneficial for designing reconfigurable photonic devices, where light can be guided or redirected without the need for physical changes to the crystal structure [14,15].

EFC analysis in nonlinear photonic crystals is crucial for understanding how different modes propagate within the crystal. In linear PCs, EFCs provide information on the allowed propagation modes and their associated group velocities. However, in nonlinear PCs with Kerr-type nonlinearity, the EFCs become intensity-dependent. As the light intensity increases, the nonlinear refractive index change alters the shape of the EFCs, potentially leading to phenomena such as bistability or multi-stability in the propagation modes. This behavior can be harnessed to develop optical limiters, modulators, and other photonic devices that operate based on the nonlinear response of the material.

Recent works on nonlinear photonic crystals have focused on ultra-fast switching, logic gate applications etc. [16-20] And also SC effect on PC have focused on the investigation of the SC properties of light in different lattice types [7,20-23] recently. In this study, we propose that Kerr-type nonlinearities in photonic crystals can enhance the self-collimation effect by broadening the operational bandwidth and improving the confinement of light within the crystal. For instance, the nonlinear interaction can flatten the EFCs further, leading to more robust self-collimation even under varying external conditions. This makes Kerr-type nonlinear PCs ideal candidates for applications in photonic circuits where tight control over light propagation is essential.

## II. Mathematical Framework

Due to the difficulty of analytical approaches, numerical methods are used for the analysis and design of photonic structures. Among them, Plane wave expansion (PWE) and Finite Difference Time Domain (FDTD) methods are dominant in terms of their performance and meet the demand for analyzing photonic crystal-based optical devices [1,24,25]. In this study, the freely available software MPB, which utilizes the plane-wave expansion (PWE) method, is employed to calculate dispersion characteristics. Additionally, the other freely available software MEEP is used for time-domain simulations. The PWE method facilitates frequency-domain analysis, while the FDTD simulations enable time-domain analysis, allowing for a comprehensive investigation of the self-collimation effect.

## A. Kerr type nonlinearity

Kerr-type nonlinearity describes the phenomenon where the refractive index of a material varies with the intensity $I$ of the light passing through it. This relationship is given by:

$$n(I) = n_0 + n_2 I$$

where $n_0$ is the linear refractive index., $n_2$ is the Kerr coefficient,, $I$ is the intensity of the light. The change in the effective refractive index modifies the photonic band structure, thus affecting the propagation of light. In a medium with Kerr-type nonlinearity, the propagation of an optical field $\mathbf{E}(z,t)$ can be described by the Nonlinear Schrödinger Equation (NLSE):

$$\frac{\partial \mathbf{E}(z,t)}{\partial z} + \frac{i\beta_2}{2}\frac{\partial^2 \mathbf{E}(z,t)}{\partial t^2} = i\gamma |\mathbf{E}(z,t)|^2 \mathbf{E}(z,t)$$

where $z$ is the propagation distance, t is time, $\beta_2$ is the group velocity dispersion (GVD) parameter, $\gamma$ is the nonlinear parameter, given by:

$$\gamma = \frac{2\pi n_2}{\lambda A_{eff}}$$

where $\lambda$ is is the wavelength and $A_{eff}$ is the effective mode area.

## B. Self-Collimation in Photonic Crystals

Self-collimation in photonic crystals refers to the ability of the crystal to guide light beams without divergence, typically arising from specific conditions in the photonic band structure. The key to understanding self-collimation lies in analyzing the equifrequency contours (EFCs) in the reciprocal lattice. The EFC is defined as:

$$\omega(k) = constant$$

where: $\omega(\mathbf{k})$ is the frequency as a function of the wave vector $\mathbf{k}$. $\mathbf{k}$ is the wave vector in the photonic crystal. For self-collimation to occur, the EFCs must be flat in certain directions. In such cases, the group velocity $\mathbf{v}_g$ which is given by:

$$v_g = \nabla_k \omega(\mathbf{k})$$

remains constant across the EFC, leading to non-divergent beam propagation. The photonic band structure can be described by solving Maxwell's equations in a periodic dielectric medium:

$$\nabla \times \left(\frac{1}{\epsilon(\mathbf{r})}\nabla \times \mathbf{H}(\mathbf{r})\right) = \left(\frac{\omega}{c}\right)^2 \mathbf{H}(\mathbf{r})$$

Where $\epsilon(\mathbf{r})$ is the spatially varying dielectric function and $\mathbf{H}(\mathbf{r})$ is the magnetic field. For a specific direction, say along the *x*-axis, the condition for self-collimation is that the curvature of the EFC is close to zero:

$$\frac{\partial^2 \omega}{\partial k_x^2} \approx 0$$

If this condition is met over a range of frequencies, light beams can propagate through the photonic crystal without spreading.

## III. SIMULATION AND RESULTS

In this study, the hexagonal lattice photonic crystal was made low-symmetric by adding two perturbation rods. All rods were selected as Kerr-type nonlinear $LiNbO_3$ (n=2.24). The radius of the rods and auxiliary perturbed rods was determined as r=0.2a, and rp=0.1a respectively. a is the lattice constant as shown in Fig. 1. Distance of perturbation rod d=0.35a. All units are taken normalized in the simulation. EFC, GVD and TOD diagrams of the structure were obtained for different NL index changes. EFC is dramatically affected by index change depicted in Fig. 2. If we look at the self-collimation properties of the normalized frequency 0.744, one can see that the SC of the wave is improved by increasing a nonlinear index change as seen in Fig. 2-3.

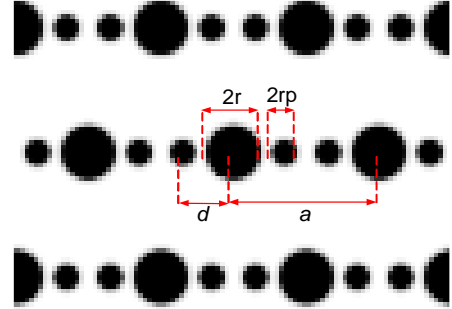

Fig. 1. Hexagonal lattice low symmetry photonic crystal structure

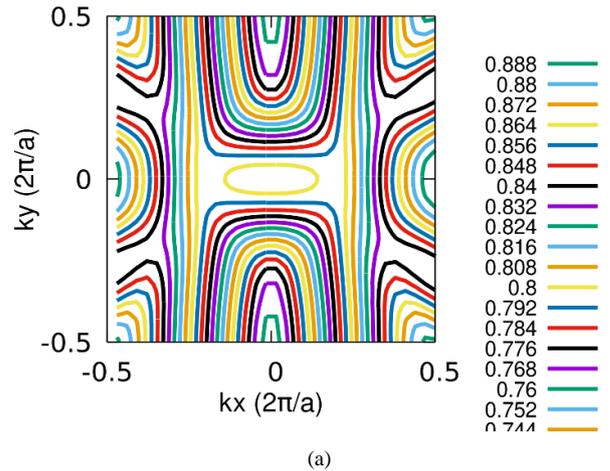

(a)

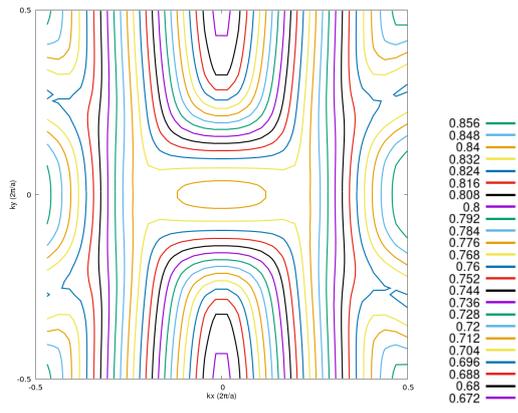

(b)

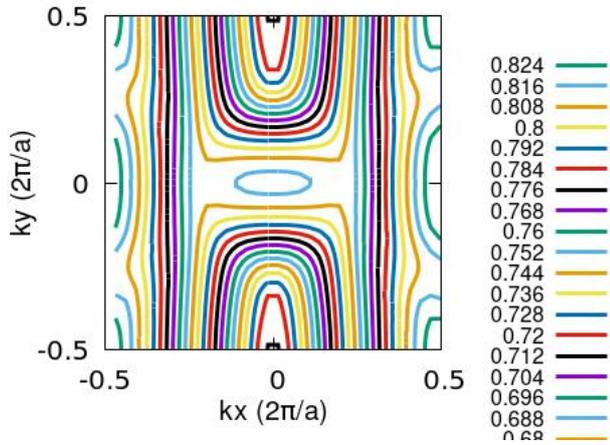

(c)

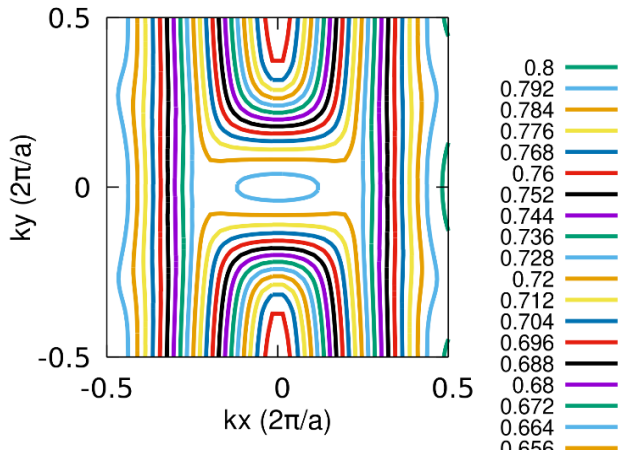

(d)

Fig. 2. EFC plots of the low symmetry hexagonal photonic crystal structure for nonlinear index change of (a-d) Δn =0 to 0.3 respectively.

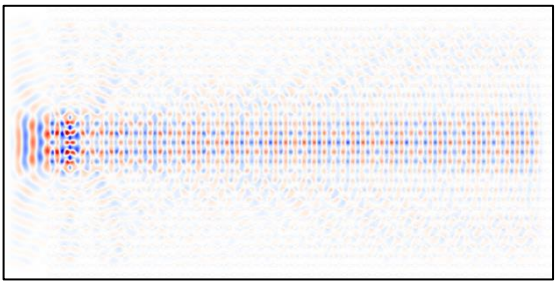

(a)

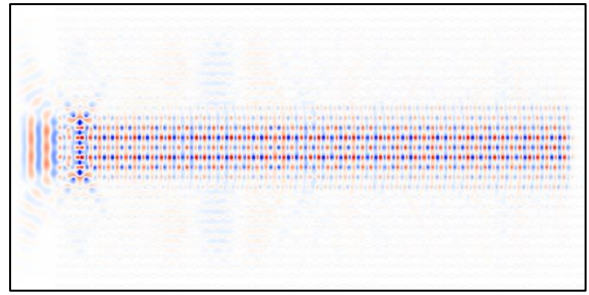

(b)

Fig. 3. Field propagation of 0.744 frequency with and without nonlinear index change

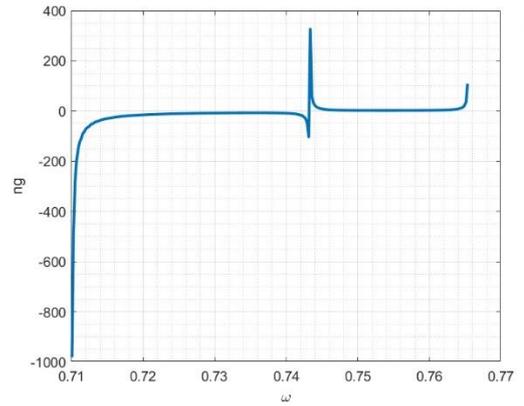

(a)

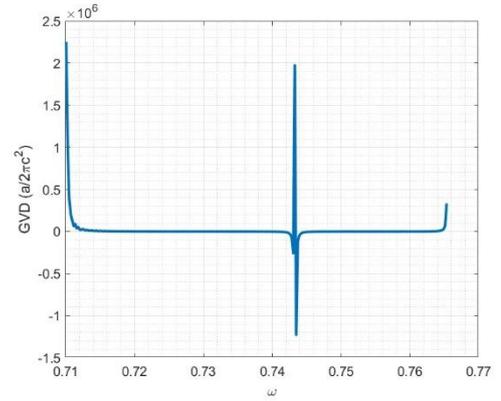

(b)

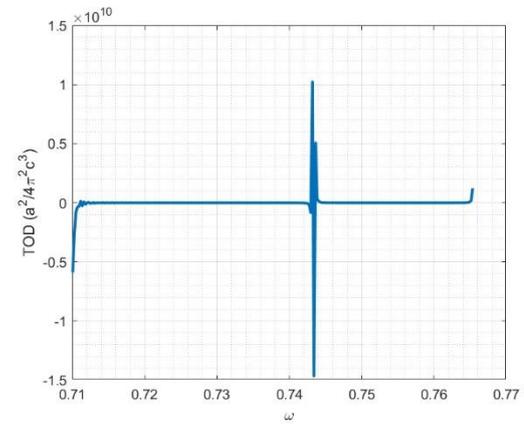

(c)

Fig. 4. a) Group index  b) GVD and c) TOD characteristics of the structure

Additionally, group velocity dispersion (GVD), third-order dispersion (TOD), and the relationship between the NL properties of PCs are presented in Fig. 4. As a result, the integration of Kerr-type nonlinearity with photonic crystals enhances the self-collimation effect through the modulation of EFCs. This capability not only opens up new avenues for dynamic photonic devices but also offers a deeper understanding of light behavior in complex nonlinear media. As research progresses, the potential for these materials in practical applications, such as optical computing and telecommunications, will continue to grow (SpringerLink) (Nature).

ACKNOWLEDGMENT

This work was supported by the Scientific and Technological Research Council of Turkey (TUBITAK) under Project No. 118E954.